\documentclass[prd,showpacs]{revtex4} 

\begin{document}

\title{The Schwarzschild solution in a Kaluza-Klein theory with two times}

\author{J. Koci\'nski  and M. Wierzbicki}
\affiliation{Faculty of Physics, Warsaw University of Technology\\
Koszykowa 75, 00-662 Warszawa, Poland}


\begin{abstract}
\noindent
A new spherically-symmetric solution is determined in a noncompactified
Kaluza-Klein theory  in which a time character is ascribed to the fifth
coordinate. This solution contains two
independent parameters which are related with mass and electric charge.
The solution exhibits a Schwarzschild radius
and represents a generalization of the Schwarzschild solution in four dimensions.
The parameter of the solution connected with the electric charge depends on
the derivative  of the fifth (second time) coordinate with respect to the
ordinary time coordinate. It is shown that the perihelic motion in
four-dimensional relativity has a counterpart in five dimensions in the
perinucleic motion of a negatively-charged particle.
If the quantization conditions of the older quantum theory are applied to that motion,
an analogue of the fine-structure formula of atomic spectra is obtained.
\end{abstract}

\maketitle

\section{INTRODUCTION}

The idea of two times in a physical theory gained an impetus
with the investigations of I. Bars and his coworkes \cite{Bars,Bars2,Bars3,Bars4,Bars5};
two-times physics provides a new perspective for understanding the one-time dynamics
from a higher-dimensional point of view.
From a single action formula of two-time physics, with the application of
gauge theory,  diverse one-time dynamical systems can be obtained.

In general relativity the second time  variable was introduced and discusssed
in the quality  of a universal parametric "historical time"   by Horwitz and
Piron \cite{Horwitz}, or as its generalization  by Burakovsky and Horwitz
\cite{Burakovsky}.

In the Kaluza-Klein theory three objections were raised against the timelike
signature of the  fifth coordinate
\cite{Bailin,Overduin,Appelquist,Appelquist2}. (1) When in the
five-dimensional action the integration over the fifth coordinate  is
performed, provided that all the derivatives with respect to $x^5$ are omitted
and the cylinder condition is accepted, the Maxwell action comes out with an
opposite sign to that of the Einstein action which is incorrect. (2) The
existence of tachyons follows from the accepted cylinder condition. (3) There
would appear closed time curves.

The problem of closed time curves in four-dimensional gravity  was
investigated in  a series of papers of Friedman, Thorne and their coworkers
\cite{Friedman,Thorn,Friedman2}. A particular attention was paid to the
question whether closed time curves violate the causality principle. The
present answer does not seem to be conclusive in that respect. In the case of
the five-dimensional gravity with two time variables an analogous
investigation has not been undertaken. It is an open question whether the
second objection is relevant in noncompactified Kaluza-Klein theories to which
this paper refers. The first objection would be relevant in a noncompactified
Kaluza-Klein theory if it were confirmed. The attitude towards a timelike
signature of the fifth coordinate is less restrained in \cite{Wesson} where
spacelike or timelike signature of the fifth cordinate is admitted depending
on the physical problem in question.

In Section 2 we start from the line element which formally is identical with
that  of Chodos and Detweiler \cite{Chodos}. The difference consists in the
spatial character of the fifth coordinate $x^5$ in \cite{Chodos} and the time
character  ascribed to this coordinate in the present case. We determine a
static five-dimensional spherically-symmetric solution on the basis of the
line element in which there are two time coordinates. This solution exhibits a
Schwarzschild radius and represents a generalization of the four-dimensional
Schwarzschild solution. The solution depends on four parameters of which two
are independent and can be related with gravitational mass and electric
charge. This is accomplished in Section 3 where we discuss the geodesic
equation in a nearly flat space. From the geodesic equation there follows a
linear relation between the ordinary time $t$ and the second time variable
$u$. In Section 4 we solve the problem of the perinucleic motion of a
negatively charged test particle (electron) moving in the field of the central
positively charged mass.  Including into the five-dimensional geometry the
quantization conditions of the older quantum theory \cite{Sommerfeld} we
derive  an analogue of Sommerfeld's relativistic energy-level formula
\cite{Sommerfeld}. 

\section{A SPHERICALLY-SYMMETRIC SOLUTION FOR A TWO-TIME LINE ELEMENT} 
\noindent We consider the line element in a five-dimensional Riemann space:

\begin{equation}
dS^2=\gamma_{ab}dx^adx^b
\label{eq1-1}
\end{equation}

\noindent
where $a,b=1,\ldots,5$ and $\gamma_{ab}$ denote the elements of the metric tensor.

In a flat space we introduce the Cartesian coordinates
$x^1,x^2,x^3=x,y,z$, $x^4=ct$, $x^5=cu$
where $c$ denotes the speed of light in the vacuum while $t$ and $u$
are expressed in the units of time. The non-zero components of the metric
tensor are: $\gamma_{11}=\gamma_{22}=\gamma_{33}=-1$, $\gamma_{44}=1$ and
$\gamma_{55}=1$ since time character is ascribed to the fifth coordinate.

With

\begin{equation}
(x^1,x^2,x^3,x^4,x^5)=(r,\theta,\varphi,ct,cu)
\label{eq001}
\end{equation}

\noindent
the spherically-symmetric line element in a flat space has the form

\begin{equation}
dS^2=-dr^2-r^2(d\theta^2+\sin^2\!\theta\, d\varphi^2)+c^2dt^2+c^2du^2
\label{eq1}
\end{equation}

\noindent
A general form of the spherically symmetric line
element in a curved space is the following:

\begin{eqnarray}
dS^2=A(r,t,u)\,c^2dt^2+B(r,t,u)\,dr^2+C(r,t,u)\,cdrdt+
\nonumber\\
+D(r,t,u)\,(d\theta^2+\sin^2\!\theta\, d\varphi^2)+E(r,t,u)\,c^2du^2
\nonumber\\
+F(r,t,u)\,cdrdu+G(r,t,u)\,c^2dudt \hskip 0.145\textwidth
\label{eq2}
\end{eqnarray}

\noindent
In an appropriate coordinate system $r',t',u'$ we can assume that 

\begin{equation}
C(r',t',u')=F(r',t',u')=0
\label{eq3}
\end{equation}

\noindent
and

\begin{equation}
D(r',t',u')=-r'^2
\label{eq4}
\end{equation}

\noindent
In the following we shall omit the "prime" of the new  coordinates $r'$, $t'$
and $u'$ and we assume that the functions $A$, $B$, $E$, and $G$ in Eq.~(4)
are of the form:

\begin{eqnarray}
A(r,t,u)=e^{\nu(r)} \qquad B(r,t,u)=-e^{\lambda(r)}\nonumber
\\
E(r,t,u)=e^{\mu(r)} \qquad G(r,t,u)=\sigma(r)
\label{eq5}
\end{eqnarray}

\noindent
with $\mu,\nu,\lambda,\sigma\to 0$ when $r\to\infty$.
The line element in Eq.~(\ref{eq2}) now takes the form:

\begin{equation}
dS^2=-e^\lambda dr^2-r^2(d\theta^2+\sin^2\theta\, d\varphi^2)+
e^\nu c^2dt^2+e^\mu c^2 du^2+\sigma c^2 dudt
\label{eq5b}
\end{equation}

\noindent
With the speed of light in the vacuum $c$ absorbed by the variables $t$
and $u$, the respective metric tensor components are

\begin{equation}
\begin{array}{lll}
\gamma_{11}=-e^\lambda\ , &
\gamma_{22}=-r^2\ , &
\gamma_{33}=-r^2\sin^2\!\theta\ ,\\
\gamma_{44}=e^{\nu}\ , &
\gamma_{55}=e^\mu\ , &
\gamma_{45}=\gamma_{54}=\sigma 
\end{array}
\label{eq6}
\end{equation}

\noindent
The determinant $\gamma$ of this metric tensor is given by

\begin{equation}
\gamma=r^4\,e^\lambda\,\sin^2\!\theta\,\big(\sigma^2-e^{\mu+\nu}\big)
\label{eq7}
\end{equation}

\noindent
The line element in Eq.~(\ref{eq5b}) is analogous to that considered by Chodos and Detweiler in Eq.~(17) of \cite{Chodos},
however, in that paper the fifth coordinate
is a space coordinate, while we assign a time character to the fifth
coordinate.
Consequently, the spherically symmetric solution connected with the line element in Eq.~(\ref{eq5b})
will be of a different form that in \cite{Chodos}.

When all derivatives with respect to the times $t$ and $u$ are
omitted, denoting by a "prime", the derivative $d/dr$ we obtain the following non-zero components of the contracted Riemann tensor

\begin{eqnarray}
R_{11}=
{\frac{1}{4\,r\,{{( {\sigma^2} 
          -\,{e^{\mu  + \nu }}  ) }^2}}}
\bigg\{
2\,r\, \left( -{\sigma^2} -\, {e^{\mu  + \nu }}
   \right) \,{{\sigma'}^{\,2}} 
\nonumber
\\
- \left( {\sigma^2} -\, {e^{\mu  + \nu }}
       \right) \,\lambda '\,
    \left( 4\,{\sigma^2} -\,4\,{e^{\mu  + \nu }}
      -{e^{\mu  + \nu }}\,r\,\mu ' -\,
      {e^{\mu  + \nu }}\,r\,\nu ' \right)
\nonumber
\\
-2\,\sigma\,r\,\sigma'\,\left[ \left( {\sigma^2} 
      -\, {e^{\mu  + \nu }}  \right) \,\lambda ' -\,
     2\,{e^{\mu  + \nu }} \,
     \left( \mu ' + \nu ' \right)  \right]
\nonumber
\\
+r\bigg[
e^{\mu+\nu}(e^{\mu+\nu}-2\sigma^2)(\mu'^2+\nu'^2)
-2\,
  {e^{\mu  + \nu }}\,{\sigma^2} \,\mu '\,\nu ' 
\nonumber
\\
+4\,{\sigma^3}\,\sigma'' -\, 4\,{e^{\mu  + \nu }}\,\sigma \,
   \sigma'' +2e^{\mu+\nu}(e^{\mu+\nu}-\sigma^2)(\mu''+\nu'')
\bigg]
\bigg\}
\end{eqnarray}

\begin{eqnarray}
R_{22}=
-{\frac{e^{-\lambda}}{2\,
     \left( {\sigma^2} -2 {e^{\mu  + \nu }}\,\right)
     }}
\bigg[
-2\,\sigma\,r\,\sigma' + {\sigma^2}\,\left( -2 + 2\,{e^{\lambda }} + 
     r\,\lambda ' \right)
\nonumber
\\
-\,{e^{\mu  + \nu }} \,
  \left( -2 + 2\,{e^{\lambda }} + r\,\lambda ' - r\,\mu ' - 
    r\,\nu ' \right)
\bigg]
\end{eqnarray}

\begin{equation}
R_{33}=R_{22}\,\sin^2\!\theta
\end{equation}

\begin{eqnarray}
R_{44}=
{\frac{-{e^{-\lambda  + \nu }}}
   {4\,r\,\left( {\sigma^2} -\, 
       {e^{\mu  + \nu }}  \right) }}
\bigg[
2\,r\,{{\sigma'}^2} - 2\,\sigma\,r\,\sigma'\,\nu '
\nonumber
\\
+\left[ 4\,{\sigma^2} - 4\,\,{e^{\mu  + \nu }}  -
    {\sigma^2}\,r\,\lambda ' +\, 
    {e^{\mu  + \nu }}\,r \,(\lambda '-\mu')  \right] \,
  \nu '
\nonumber
\\
+r\,\left( 2\,{\sigma^2} -\, {e^{\mu  + \nu }} \right)
     \,{{\nu '}^2} + 2\,r\,
   \left( {\sigma^2} -\, {e^{\mu  + \nu }}  \right) \,
   \nu ''
\bigg]
\end{eqnarray}

\begin{eqnarray}
R_{45}=R_{54}=
{\frac{1}{4\,{e^{\lambda }}\,r\,
     \left( {\sigma^2} -\, {e^{\mu  + \nu }}  \right)
     }}
\bigg[
{\sigma^2}\,\sigma'\,\left( -4 + r\,\lambda ' \right)
\nonumber
\\
-\,
{e^{\mu  + \nu }} \,\sigma'\,
  \left( -4 + r\,\lambda ' + r\,\mu ' + r\,\nu ' \right)
\nonumber
\\
+2\,r\,\left( {e^{\mu  + \nu }}\,\sigma \,\mu '\,
     \nu ' - {\sigma^2}\,\sigma'' +
    {e^{\mu  + \nu }} \,\sigma'' \right)
\bigg]
\end{eqnarray}

\begin{eqnarray}
R_{55}=
{\frac{-{e^{-\lambda  + \mu }} }
   {4\,r\,\left( {\sigma^2} -\,
       {e^{\mu  + \nu }}  \right) }}
\bigg[
2\,r\,{{\sigma'}^2} - 2\,\sigma\,r\,\sigma'\,\mu ' +
  r\,\left( 2\,{\sigma^2}-\,{e^{\mu  + \nu }}
      \right) \,{{\mu '}^2}
\nonumber
\\
+\mu '\,\left( 4\,{\sigma^2} - 
    4\,{e^{\mu  + \nu }}  -
    {\sigma^2}\,r\,\lambda ' + 
    {e^{\mu  + \nu }}\,r \,\lambda ' -
    {e^{\mu  + \nu }}\,r \,\nu ' \right)
\nonumber
\\
+2\,r\,\left( {\sigma^2} -\,{e^{\mu  + \nu }} \right)
    \,\mu ''
\bigg]
\end{eqnarray}

\noindent
These are equated to zero, yielding five equations
for the four unknown functions $\mu,\nu,\lambda,\sigma$ in Eq.~(\ref{eq5}).

It can be verified that a
solution of the equations

\begin{equation}
R_{11}=R_{22}=R_{44}=R_{55}=R_{45}=0
\label{eq9-0}
\end{equation}

\noindent
is given by the following functions:

\begin{equation}
e^\nu=1-\frac{{\cal G}}{r}\,,\quad
e^\mu=1-\frac{{\cal C}}{r}\,,\quad
e^\lambda=\Big(1-\frac{{\cal R}}{r}\Big)^{-1}\,,\quad
\sigma=\frac{{\cal P}}{r}
\label{eq9}
\end{equation}

\noindent
where the real parameters ${\cal G},{\cal C},{\cal R}$ and ${\cal P}$ satisfy the conditions

\begin{equation}
{\cal R}={\cal G}+{\cal C}\,,\quad {\cal P}^2={\cal G}{\cal C}
\label{eq10}
\end{equation}

\noindent
The parameters ${\cal G}$ and ${\cal C}$ will be related  with gravitational 
mass and electric charge, respectively, on the basis of the linearized
form of the geodesic equation.


\section{THE PARAMETERS IN THE SPHERICALLY-SYMMETRIC SOLUTION}

We consider the two-times-independent metric tensor of the form:

\begin{equation}
\gamma_{ab}=\gamma_{ab}^{(5)}+\eta_{ab}\,,\quad a,b=1,\ldots,5
\label{eq11}
\end{equation}

\noindent 
where $\gamma_{ab}^{(5)}$ is the five-dimensional flat-space metric tensor specified after
Eq.~(\ref{eq1-1}), and $\eta_{ab}$
represents a small perturbation, due to the presence of a gravitating body with an electric
charge. The perturbation vanishes very far from the body.
To show that the $\eta_{ab}$ terms are the agents of gravitational and electrostatic forces
we consider the geodesic equation of motion in the Riemann space with the above metric.
We assume that the velocity of a test particle (with mass and electric charge)
along the geodesic line is much smaller than the speed of light $c$.
Using the nearly flat-space metric tensor in Eq.~(\ref{eq11}) we then obtain the line element

\begin{equation}
dS^2=-(dx^1)^2-(dx^2)^2-(dx^3)^2+(dx^4)^2+ (dx^5)^2+\eta_{ab}dx^{a}dx^{b}
\label{eq12}
\end{equation}

\noindent
from which we obtain:

\begin{equation}
\Big(\frac{dS}{dt}\Big)^2=c^2
\Big[
1-\frac{v^2}{c^2}+\Big(\frac{du}{dt}\Big)^2+\frac{1}{c^2}
\kern 0.1em\eta_{ab}\frac{dx^{a}}{dt}\frac{dx^{b}}{dt}
\Big]
\label{eq13}
\end{equation}

\noindent
where $v^j=dx^j/dt$, $j=1,2,3$. 
Retaining in Eq.~(\ref{eq13}) the terms of the first order in $v/c$  we obtain:

\begin{equation}
\Big(\frac{dS}{dt}\Big)^2\approx
c^2
\Big[
1+\Big(\frac{du}{dt}\Big)^2(1+\eta_{55})+\eta_{44}
+2\eta_{45}\frac{du}{dt}
\Big]
\label{eq14}
\end{equation}

\noindent
We had to assume that
$v^2/c^2 \ll (du/dt)^2$, since otherwise we would have to omit $(du/dt)^2$
together with $v^2/c^2$.

We next apply the same approximations to the geodesic equation

\begin{equation}
\frac{d^2x^a}{dS^2}+\Gamma_{bc}^a\frac{dx^b}{dS}\frac{dx^c}{dS}=0
\label{eq15}
\end{equation}

\noindent
From the form of the metric in Eq.~(\ref{eq12}) follows that each Christoffel symbol
linearly depends on the perturbation $\eta_{ab}$.
With the accuracy to terms of order $v/c$ we obtain the equality

\begin{equation}
\Gamma_{bc}^a\frac{dx^b}{dS}\frac{dx^c}{dS}=
c^2
\Big[
\Gamma^a_{44}+2
\frac{du}{dt}\Gamma^a_{45}+\Big(
\frac{du}{dt}\Big)^2\Gamma^a_{55}
\Big]
\Big(
\frac{dt}{dS}\Big)^2
\label{eq17}
\end{equation}

\noindent
From Eq.~(\ref{eq15}) and (\ref{eq17}) follows the equation

\begin{equation}
\frac{d^2x^a}{dt^2}=
c^2
\Big[
\Gamma^a_{44}+2
\frac{du}{dt}\Gamma^a_{45}+\Big(
\frac{du}{dt}\Big)^2\Gamma^a_{55}
\Big]
\label{eq19}
\end{equation}

\noindent
Since $\eta_{ab}$ are independent of $t$ and $u$ the Christoffel symbols
in Eq.~(\ref{eq19}) vanish for $a=4,5$. For $a=1,2,3$ we have

\begin{equation}
\Gamma_{44}^a=\frac{1}{2}\frac{\partial\eta_{44}}{\partial x^a}\,,\ \ 
\Gamma_{45}^a=\frac{1}{2}\frac{\partial\eta_{45}}{\partial x^a}\,,\ \ 
\Gamma_{55}^a=\frac{-1}{2}\frac{\partial\eta_{55}}{\partial x^a}
\label{eq20}
\end{equation}

\noindent
and from Eq.~(\ref{eq19}), for $a=5$ we find that
$d^2u/dt^2=0$, hence

\begin{equation}
u=wt+u_0
\label{eq21}
\end{equation}

\noindent
where $w$ and $u_0$ are constants. With $u_0=0$, $w>0$, considering Eqs.~(\ref{eq20}),
and with $w=du/dt$ we obtain from Eq.~(\ref{eq19}) the equation

\begin{equation}
\frac{d^2x^a}{dt^2}=-
\frac{c^2}{2}
\Big[
\frac{\partial\eta_{44}}{\partial x^a}+2w
\frac{\partial\eta_{45}}{\partial x^a}+w^2\frac{\partial\eta_{55}}{\partial x^a}
\Big]
\label{eq22}
\end{equation}

\noindent
On the basis of this equation we will determine the parameters
${\cal G}$ and ${\cal C}$ in Eqs.~(\ref{eq9}) and (\ref{eq10}).

We assume that the test particle is an electron with mass $m$. 
Multiplying both sides of Eq.~(\ref{eq22}) with $m$, we can identify the first
term  on the r.h.s.  with the gravitational force acting on the electron mass
$m$, and to one of the two remaining terms we can ascribe the meaning of the
Coulomb force acting on the electron charge $e$. It will appear that we have
to relate  the third term on the r.h.s. of Eq.~(\ref{eq22}) with the Coulomb
force.

We begin with the first term on the r.h.s. of Eq.~(\ref{eq22}) and write:

\begin{equation}
\Big(m\frac{d^2x^a}{dt^2}\Big)_{\mbox{\small mechanical}}=-m\frac{c^2}{2}\frac{\partial \eta_{44}}{\partial x^a}=
-m\frac{\partial\psi}{\partial x^a}
\label{eq22a}
\end{equation}

\noindent
with $\psi$ denoting the gravitational potential of the mass $M$,  where
$\kappa$ is the gravitational constant,

\begin{equation}
\psi=-\kappa\frac{M}{r}
\label{eq23}
\end{equation}

\noindent
thus obtaining the equality

\begin{equation}
\eta_{44}=\frac{2}{c^2}\psi
\label{eq24}
\end{equation}

\noindent
and from this and from Eq.~(\ref{eq12}), the equality

\begin{equation}
\gamma_{44}=1+\eta_{44}
\label{eq24a}
\end{equation}

\noindent
From the first of Eqs.~(\ref{eq9}) and from Eqs.~(\ref{eq23}), (\ref{eq24})
and  (\ref{eq24a}) we then find that

\begin{equation}
{\cal G}=\frac{2\kappa M}{c^2}
\label{eq25}
\end{equation}

\noindent
as in the case of the Schwarzschild solution in 4 dimensions.

We next consider the third term on the r.h.s of Eq.~(\ref{eq22}). Let
$M$ denote the proton mass and
let $\varphi$ denote the electrostatic potential of the proton charge $Q$

\begin{equation}
\varphi=\frac{Q}{4\pi\varepsilon_0 r}
\label{eq26}
\end{equation}

\noindent
where $_0$ denotes the vacuum electric permeability.  On the basis
of Eq.~(\ref{eq22}) we write 

\begin{equation}
\Big(m\frac{d^2x^a}{dt^2}\Big)_{\mbox{\small electrical}}=-\frac{1}{2} m c^2
w^2  \frac{\partial \eta_{55}}{\partial x^a} =-e\frac{\partial
\varphi}{\partial x^a} \label{eq27}
\end{equation}

\noindent
for the electrostatic force acting on the electric charge $e$ connected with the mass $m$.
From Eq.~(\ref{eq27}) and (\ref{eq28}) we obtain

\begin{equation}
\frac{1}{2} c^2 w^2 \eta_{55}=\frac{e}{m}\varphi=\frac{Qe}{4\pi\varepsilon_0
rm} \label{eq28}
\end{equation}

\noindent
and hence

\begin{equation}
\eta_{55}=\frac{2eQ}{4\pi\varepsilon_0 mc^2w^2r} 
\label{eq28a}
\end{equation}

\noindent
On the other hand from the second of Eqs.~(\ref{eq9}) and from Eq.~(\ref{eq12}) we obtain

\begin{equation}
\gamma_{55}=1+\eta_{55}=1-\frac{{\cal C}}{r}
\label{eq29}
\end{equation}

From Eqs.~(\ref{eq28a}) and (\ref{eq29}) we find that

\begin{equation}
{\cal C}=\frac{-2eQ}{4\pi\varepsilon_0 mc^2w^2} 
\label{eq30}
\end{equation}

\noindent
From the second of Eqs.~(\ref{eq10}) and from Eq.~(\ref{eq26}) follows that
${\cal C}$  must be positive. This implies that when the charge $Q$ is
assumed positive, the charge $e$ must be negative and vice versa.
When $e$ and $Q$ are of the same sign the second condition in Eq.~(\ref{eq10}) cannot be
fulfilled. This formula depends on the parameter $w^2=(du/dt)^2$, which in
connection with Eq.~(\ref{eq14}) has to be much larger than $v^2/c^2$.
 
We now can determine the parameter ${\cal P}$ in Eq.~(\ref{eq10}).
Owing to Eq.~(\ref{eq30}) we obtain

\begin{equation}
{\cal P}=\sqrt{{\cal G}{\cal C}}=\frac{1}{wc^2}\sqrt{\frac{\kappa M|eQ|}{\pi\varepsilon_0 m}}
\label{eq30c}
\end{equation}

\noindent
where $-eQ$ in Eq.~(\ref{eq30}) has been replaced by $|eQ|$, since $e$ and $Q$
must have opposite signs. With the gravitational constant $\kappa=6.673\times
10^{-11}\,\mbox{N}\cdot\mbox{M}^2\cdot\mbox{K}$, the vacuum electric
permeability $\varepsilon_0=0.885\times
10^{-11}\,\mbox{Q}\cdot\mbox{V}^{-1}\cdot\mbox{M}^{-1}$, with $e/m=1.76\times
10^{11}\,\mbox{Q}\cdot\mbox{K}$, and the absolute value of electron charge
$|e|=1.6\times 10^{-19}\,\mbox{Q}$, where N=newton, M=meter, K=kilogram,
Q=coulomb, V=volt \cite{Sommerfeld2} identifying the mass $M$ in Eq.
(\ref{eq25}) with the proton mass  we find from Eq.~(\ref{eq25}) that 

\begin{equation} 
{\cal G}\approx 2.4\times 10^{-54}\,\mbox{M} 
\label{eq31}
\end{equation}

\noindent
With $Q$ denoting the proton charge, from Eq.~(\ref{eq30}) we find that 

\begin{equation}
{\cal C}\approx 5.6\,w^{-2} \times 10^{-15}\,\mbox{M}
\label{eq31a}
\end{equation}

\noindent
and then with $Q=|e|$ from Eq.~(\ref{eq30c}) we obtain

\begin{equation}
{\cal P}\approx 1.2\,w^{-1}\times 10^{-34}\,\mbox{M}
\label{eq31b}
\end{equation}

We now can answer the question for the Schwarzschild radius.  
From Eqs.~(\ref{eq10}), (\ref{eq25}) and (\ref{eq30}) we find that in Eq.~(19) 
\begin{equation}
{\cal R}={\cal G}+{\cal C}= 2.4\times 10^{-54}\,\mbox{M}+5.6\,w^{-2}\times
10^{-15}\,\mbox{M} \label{eq31c}
\end{equation}

\noindent
If $w^{-2}$ is not extremely small, the Schwarzschild radius is determined
by the parameter ${\cal C}$ connected with the electric charge of the proton.

We now explain why it is not possible to relate the term ${\cal P}/ r$ with
the Coulomb potential. If  ${\cal P}/ r$ were related with the Coulomb
potential, from Eq.~(\ref{eq22}) we then would obtain: 
\begin{equation}
\Big(m\frac{d^2x^a}{dt^2}\Big)_{\mbox{\small electrical}}=-mc^2w
\frac{\partial \eta_{45}}{\partial x^a}= -e\frac{\partial \varphi}{\partial
x^a} \label{eq32}
\end{equation}

\noindent
with $\varphi$ defined in Eq.~(\ref{eq26}). From Eqs.~(\ref{eq26}) and
(\ref{eq32}) we would obtain: 
\begin{equation}
\eta_{45}=\frac{eQ}{4\pi\varepsilon_0 r m c^2 w}
\label{eq33}
\end{equation}

\noindent
and since from Eqs.~(\ref{eq5b}) and (\ref{eq12}) we have

\begin{equation}
\gamma_{45}=\eta_{45}=\frac{{\cal P}}{r}
\label{eq34}
\end{equation}

\noindent
we would find 

\begin{equation}
{\cal P}=\frac{eQ}{4\pi\varepsilon_0 m c^2 w}
\label{eq35}
\end{equation}

\noindent
The parameter ${\cal C}$ now is obtained from Eq.~(\ref{eq10}) 

\begin{equation}
{\cal C}={\cal P}^2{\cal G}^{-1}
\label{eq35b}
\end{equation}

\noindent
The parameter ${\cal R}$ is determined in Eq.~(\ref{eq10}).
Inserting into Eq.~(\ref{eq35b}) the numerical values for $\cal P$ and $\cal
G$  given in Eqs.~(\ref{eq31b}) and (\ref{eq31})  we find that
${\cal C}\approx 3\,w^{-2}\times 10^{24}\,\mbox{M}$, hence, unless $w^{-2}$ 
is extremely small, the respective ${\cal R}={\cal G}+{\cal C}$ is unacceptable
as a candidate for the Schwarzschild radius of the proton. This seems to
indicate that we cannot identify the force  connected with $\eta_{45}$ with
the Coulomb force between the central charge $Q$ and the charge $e$ of a test
particle.


\section{THE RELATIVISTIC ENERGY-LEVEL FORMULA}

We set up the variational problem for the square of the interval 
in Eq.~(\ref{eq5b}) (see for example \cite{Adler}) in the form

\begin{eqnarray}
\delta \int\Big[
\Big(1-\frac{{\cal G}}{r}\Big)
c^2\,\dot t^2 +
\Big(1-\frac{{\cal C}}{r}\Big)c^2\,\dot u^2+
\frac{\sqrt{{\cal GC}}}{r}c^2\,\dot t \dot u -
\nonumber\\
\Big(1-\frac{{\cal G}+{\cal C}}{r}\Big)^{-1} \dot r^2 -
r^2(\dot\theta^2+\dot\varphi^2\,\sin^2\theta)
\Big]dS=0
\label{eq36}
\end{eqnarray}

\noindent
where the "dot" denotes $d/dS$.
The Euler-Lagrange equations for $\theta$, $\varphi$, $t$ and $u$
yield the equalities:

\begin{equation}
\frac{d}{dS}(r^2\,\dot\theta)-r^2 \dot\varphi^2\sin\theta\cos\theta=0
\label{eq37}
\end{equation}

\begin{equation}
\frac{d}{dS}(r^2\dot\varphi\sin^2\theta)=0
\label{eq38}
\end{equation}

\begin{equation}
\frac{d}{dS}\Big[
2\Big(1-\frac{{\cal G}}{r}\Big)\dot t+\frac{\sqrt{{\cal GC}}}{r}\dot u
\Big]=0
\label{eq39}
\end{equation}

\begin{equation}
\frac{d}{dS}\Big[
2\Big(1-\frac{{\cal C}}{r}\Big)\dot u+\frac{\sqrt{{\cal GC}}}{r}\dot t
\Big]=0
\label{eq40}
\end{equation}

\noindent
Dividing the expression for the interval in Eq.~(\ref{eq5b}) by $dS^2$ we obtain the equation 
for~$\dot r$

\begin{eqnarray}
1=
\Big(1-\frac{{\cal G}}{r}\Big)
c^2\dot t^2 +
\Big(1-\frac{{\cal C}}{r}\Big)c^2\dot u^2+
\frac{\sqrt{{\cal GC}}}{r}c^2\dot t \dot u -
\nonumber\\
\hskip 5em
\Big(1-\frac{{\cal G}+{\cal C}}{r}\Big)^{-1} \dot r^2 
-r^2(\dot\theta^2+ \dot\varphi^2\sin^2\theta)
\label{eq41}
\end{eqnarray}

We assign the charge $Z|e|$ to the central mass
and will determine the perinucleic motion of an electron in the Coulomb field.
We observe that the term with
$\sqrt{{\cal GC}}$ determines the main influence of the gravitational field,
connected with the central singularity, on the energy levels. 
This influence is very small in comparison with the influence of the Coulomb
field represented by the parameter ${\cal C}$. In Eqs.~(\ref{eq39}), 
(\ref{eq40}), and (\ref{eq41}) we omit the terms containing the factors ${\cal
G}/r$ or $\sqrt{{\cal GC}}/r$, since they are small in comparison with the
terms containing ${\cal C}/r$, thus obtaining from Eq.~(\ref{eq41}) the
equation  

\begin{equation} 
1=
c^2\dot t^2 +
\Big(1-\frac{{\cal C}}{r}\Big)c^2\dot u^2-
\Big(1-\frac{{\cal C}}{r}\Big)^{-1} \dot r^2 -
r^2(\dot\theta^2+ \dot\varphi^2\sin^2\theta)
\label{eq41a}
\end{equation}

By an appropriate orientation of the coordinate axes we can make
$\theta=\pi/2$  and $d\theta/dS=\dot\theta=0$, for some initial value of
$S$. From Eq.~(\ref{eq37}) then follows that for all values of $S$  we have
$\theta=\pi/2$. Substituting this value of $\theta$ into Eq.~(\ref{eq38}) we
obtain  

\begin{equation}
r^2\frac{d\varphi}{dS}=k=\mbox{const}
\label{eq42}
\end{equation}

\noindent
while from Eqs.~(\ref{eq39}) and (\ref{eq40}) we obtain 

\begin{equation}
\frac{dt}{dS}=\tau=\mbox{const}
\label{eq43}
\end{equation}

\noindent
and

\begin{equation}
\frac{du}{dS}\Big(1-\frac{{\cal C}}{r}
\Big)=\rho=\mbox{const}
\label{eq44}
\end{equation}

\noindent
Substituting the expressions in Eqs.~(\ref{eq42}), (\ref{eq43})  and
(\ref{eq44}) into Eq.~(\ref{eq41a}) we obtain the equation for $r=r(S)$

\begin{equation}
\Big(\frac{dr}{dS}
\Big)^2=
[c^2(\tau^2+ \rho^2)-1]+(1-c^2\tau^2)\frac{{\cal C}}{r}-\frac{k^2}{r^2}+
\frac{{\cal C}k^2}{r^3} 
\label{eq45}
\end{equation}

\noindent
From this equation in the new variable $v=1/r$, in  the customary way
\cite{Adler} we obtain the equation 

\begin{equation}
v^{\prime\prime}+v={\cal A}+\frac{\beta}{{\cal A}}v^2
\label{eq46}
\end{equation}

\noindent
where 

\begin{equation}
{\cal A}=\frac{(1-c^2\tau^2)\,{\cal C}}{2k^2} \qquad\mbox{and}  \qquad
\beta=\frac{3}{2}{\cal A}\,{\cal C} \label{eq47}
\end{equation}

\noindent
Eq.~(\ref{eq46}) determines the perinucleic motion.

We now intend to determine a formula for the energy levels of the test particle. 
From Eq.~(\ref{eq45}) we obtain the expression

\begin{eqnarray}
\frac{dr}{dS}=\frac{dr}{dt}\frac{dt}{dS}=
\tau\frac{dr}{dt}=
\nonumber\\
\hskip 3em
\sqrt{
[c^2(\tau^2+ \rho^2)-1]+(1-c^2\tau^2)\frac{{\cal C}}{r}-\frac{k^2}{r^2}+
\frac{{\cal C}k^2}{r^3} }
\label{eq48}
\end{eqnarray}

\noindent
We take over from Sommerfeld (p. 277 in \cite{Sommerfeld}) the quantization conditions

\begin{equation}
\oint m\frac{dr}{dt}dr=n_{r}h\,,\quad n_r=0,1,2,\ldots
\label{eq49}
\end{equation}

\begin{equation}
\int\limits_0^{2\pi} mr^2\frac{d\varphi}{dt}d\varphi=n_{\varphi}h\,,\quad 
n_\varphi=1,2,3,\ldots \label{eq49a}
\end{equation}

\noindent
where $m$ denotes the electron mass, and $h$ is Planck's constant. From Eq.
(\ref{eq48})   we find that 

\begin{equation}
m\frac{dr}{dt}=
\sqrt{
\frac{m^2}{\tau^2}\Big\{[c^2(\tau^2+ \rho^2)-1]+(1-c^2\tau^2) \frac{{\cal
C}}{r}-\frac{k^2}{r^2}+\frac{{\cal C}k^2}{r^3}\Big\} }
\label{eq50}
\end{equation}

\noindent
From Eqs.~(\ref{eq42}) and (\ref{eq43}) and from Eq.~(\ref{eq49a})  we find
that 

\begin{equation}
k=\frac{\tau}{m}n_{\varphi} \hbar
\label{eq51}
\end{equation}

\noindent
where $\hbar=h/2\pi$. The integral in Eq.~(\ref{eq49})  with the integrand
given in Eq.~(\ref{eq50}) was calculated in \cite{Sommerfeld}. It has the
value  

\begin{eqnarray}
I=\oint\sqrt{A_0+2\frac{A_1}{r}+\frac{A_2}{r^2}+\frac{A_3}{r^3}}dr=
\nonumber\\
\hskip 3em
-2\pi i 
\Big( \sqrt{A_2}-\frac{A_1}{\sqrt{A_0}}-\frac{A_1A_3}{2A_2\sqrt{A_2}}
\Big)
\label{eq52}
\end{eqnarray}

\noindent
where $\sqrt{A_2}$ is negative imaginary.
Comparing Eq.~(\ref{eq50}) with the integrand on the l.h.s. of Eq.
(\ref{eq52})  we find that 

\begin{eqnarray}
A_0=\frac{m^2}{\tau^2}\Big[c^2(\tau^2+ \rho^2)-1\Big]
\qquad
A_1=\frac{m^2}{2\tau^2}(1-c^2\tau^2)\,{\cal C}
\nonumber\\
A_2=-\frac{m^2k^2}{\tau^2}=-n_{\varphi}^2\hbar^2
\qquad
A_3=\frac{m^2}{\tau^2}{\cal C}k^2={\cal C}n_{\varphi}^2\hbar^2
\label{eq53}
\end{eqnarray}

\noindent
The three terms on the r.h.s. of Eq.~(\ref{eq52}) take the form

\begin{equation}
-2\pi i \sqrt{A_2}=-2\pi n_{\varphi} \hbar
\label{eq54}
\end{equation}

\begin{equation}
2\pi i \frac{A_1}{\sqrt{A_0}}=2\pi i
\frac{m(1-c^2\tau^2)\,{\cal C}}{2\tau\sqrt{c^2(\tau^2+ \rho^2)-1}}
\label{eq55}
\end{equation}

\begin{equation}
\pi i \frac{A_1A_3}{A_2\sqrt{A_2}}=2\pi \frac{(1-c^2\tau^2)\,{\cal C}^2m^2}
{4n_\varphi \hbar \tau^2} \label{eq56}
\end{equation}

\noindent
From Eqs.~(\ref{eq48}), (\ref{eq50}), (\ref{eq52}) and  (\ref{eq54})  through
(\ref{eq56}), introducing the expression for ${\cal C}$ given in
Eq.~(\ref{eq30}), with $\alpha=e^2/4\pi\varepsilon_0\hbar c$ we obtain
the equality

\begin{equation}
\frac{ i Z\alpha(1-c^2\tau^2)}{c\sqrt{c^2(\tau^2+ \rho^2)-1}}
\Big(
\frac{dt}{du}
\Big)^2=
n_r+n_\varphi-\Bigg[
\frac{1-c^2\tau^2}{c^2\tau^2}
\Big(
\frac{dt}{du}
\Big)^4
\Bigg]
\frac{\alpha^2Z^2}{n_\varphi}
\label{eq57}
\end{equation}
 
\noindent
The respective expression for the relativistic Kepler motion in  a flat space
in Eq.~(26) on p. 278 of \cite{Sommerfeld} is given by

\begin{equation}
\Big(
1+\frac{W}{m_0c^2}
\Big)^{-2}=1+\frac{\alpha^2Z^2}{\big[
n_r+\sqrt{n_\varphi^2-\alpha^2Z^2}
\big]^2}
\label{eq58}
\end{equation}

\noindent
In Sommerfeld's formula in Eq.~(\ref{eq58}) the rest energy $m_0c^2$  is not
included into the energy $W$. If on the r.h.s. of Eq.~(\ref{eq57}) we put 

\begin{equation}
\frac{1-c^2\tau^2}{c^2\tau^2(du/dt)^4}=\frac{1}{2}
\label{eq59}
\end{equation}

\noindent
then writing
$\tau^2=d^2\,c^{-2}$
with real $d$, we obtain Eq.~(\ref{eq59}) in the form
$2-2d^2=d^2(du/dt)^4$,
which implies $d^2<1$. Taking Eq.~(\ref{eq59}) into account 
we can represent the square of Eq.~(\ref{eq57}) in the form

\begin{equation}
3- \frac{2c^2\rho^2}{d^2-1}=1+\frac{Z^2\alpha^2}{\Big[n_r+n_\varphi-
\frac{\displaystyle \alpha^2Z^2}{\displaystyle 2n_\varphi}\Big]^2}
\label{eq62} \end{equation}

\noindent
The l.h.s. of this equality can be expressed through the energy $W$.  Eq.
(\ref{eq62}) then leads to the Sommerfeld formula for the energy levels of an
electron in a hydrogen-like atom, in which $\sqrt{n_\varphi^2-Z^2\alpha^2}$
is approximated by $n_\varphi(1-Z^2\alpha^2/2n_\varphi)$. 

We observe that an analogue of Eq.~(\ref{eq57}) referring to the gravitational
interaction can be obtained. We start from Eq.~(\ref{eq41}) in which the terms
depending on $\cal C$ and ${\cal G}{\cal C}$ are neglected. This can be
done since the conditions in Eq.~(\ref{eq10}) allow to put ${\cal
C}=0$ We then obtain the equation

\begin{equation}
\frac{dr}{dS}=
\sqrt{
[c^2(\tau_1^2+ \rho_1^2)-1]+(1- c^2\rho_1^2)\frac{{\cal G}}{r}-
\frac{k^2}{r^2}+\frac{{\cal G}k^2}{r^3} }
\label{eq63}
\end{equation}

\noindent
in which now we have

\begin{eqnarray}
\frac{dt}{dS}\Big(1-\frac{{\cal G}}{r}\Big)=\tau_1=\mbox{const}\\
\frac{du}{dS}=\rho_1=\mbox{const}
\label{eq64}
\end{eqnarray}

\noindent
replacing $\tau$ and $\rho$ in Eqs.~(\ref{eq43}) and (\ref{eq44}), respectively.
If on the left hand side of Eq.~(\ref{eq63}) we introduce

\begin{equation}
\frac{dr}{dS}=\frac{dr}{dt}\frac{dt}{dS}=\frac{dr}{dt}\,\frac{\tau_1}{1-{\cal G}/r}
\label{eq65}
\end{equation}

\noindent
we obtain an analogue of Eq.~(\ref{eq50}) in which under the square root 
there appear additional terms proportional to  $1/r^4$ and $1/r^5$.  The latter
terms remain also when we omit the fifth dimension and consider
four-dimensional relativity which means $\rho_1=0$ in Eq.~(\ref{eq64}). The
respective integration can be performed and its result does not lead to the
formula of the type in Eq.~(\ref{eq57}).
 
If, however, on the left hand side of Eq.~(\ref{eq63}) we write

\begin{equation}
\frac{dr}{dS}=\frac{dr}{du}\frac{du}{dS}=\rho_1\frac{dr}{du}
\label{eq66}
\end{equation}

\noindent
and apply the quantization condition in Eq.~(\ref{eq49}), with $dr/dt$ 
replaced by $dr/du$, and in Eq.~(\ref{eq49a}) replace $d\varphi/dt$ by
$d\varphi/du$ we obtain an analogue of Eq.~(\ref{eq50}) which leads to
equality 

\begin{equation}
\frac{i\,\,\alpha_g Z(1- c^2\rho_1^2)} {c\rho_1\sqrt{c^2(\tau_1^2+
\rho_1^2)-1}} =
n_r+n_\varphi-\Bigg[
\frac{1- c^2\rho_1^2}{c^2\rho_1^2}
\Bigg]
\frac{\alpha_g^2Z^2}{n_\varphi}
\label{eq67}
\end{equation}

\noindent
which has the structure of Eq.~(\ref{eq57}), where
$Zm=M$ denotes the mass at the origin of the coordinates with $m$ denoting the
mass of the electron, and where $\alpha_g=\kappa m^2/\hbar c$ is the
fine-structure  constant of the gravitational interaction \cite{Wesson}.
Although Eq.~(\ref{eq67}) does not seem to have an experimental significance
it is worthwhile to point out,  that it has been obtained with the help of
the second time variable $u$.  It cannot be obtained with the ordinary time
variable $t$ employed in the quantization conditions. 


\section{CONCLUSIONS AND DISCUSSION}

Assuming that in a non-compactified Kaluza-Klein theory the fifth coordinate $x^5$ has time
character in the five-dimensional line element,
we have determined a Schwarzschild type solution of the five-dimensional Einstein
equations in the vacuum. 
The two independent parameters of that solution have been related  with mass and electric
charge, respectively.
The solution is characterized with a Schwarzschild radius
whose magnitude is predominantly determined by the electric-charge parameter.

The perihelic motion in four-dimensional relativity has a counterpart in the perinucleic
motion of an electron in a Kaluza-Klein theory with two times.
If the quantization conditions of the older quantum theory   are included
into the five-dimensional geometry, the perinucleic motion of an electron
leads to the fine structure  of line spectra, analogous to that determined by Sommerfeld's
formula for hydrogen-like atoms.

The parameter $\cal C$ which determines the Schwarzschild radius and the parameter $\cal P$
connected with a new force depend on the derivative $du/dt$ of the second time
coordinate $u$ with respect to the ordinary time coordinate $t$. Their numerical
values therefore hinge on the magnitude of that quantity. 
The indicated physical meaning of those parameters, however, is not impaired
by the lack of knowledge of the magnitude of the quantity $du/dt$ as long as
it is not extremely large or extremely small in comparison with 1.

There exists an extensive literature concerning spherically-symmetric solutions
in the Kaluza-Klein theory.  We only name the five-dimensional spherically symmetric 
solutions determined by Chodos and Detweiler  \cite{Chodos}, by Ponce de Leon and Wesson
\cite{Ponce} and by Wesson \cite{Wesson}.
Those solutions are based on the assumption of a spatial character of the
fifth coordinate. A thorough discussion of those solutions is given in Overduin
and Wesson \cite{Overduin} and in Wesson \cite{Wesson}.

The presented results seem to contribute a new element to the discussion of two-time physics consequences.

\end{document}